\documentclass[conference]{IEEEtran}
\IEEEoverridecommandlockouts
\usepackage{cite}
\usepackage{amsmath,amssymb,amsfonts}
\usepackage{algorithmic}
\usepackage[pdftex]{graphicx}
\usepackage{textcomp}
\usepackage{caption}
\usepackage{subcaption}
\usepackage{booktabs}
\usepackage[dvipsnames]{xcolor}
\usepackage{tikz}
\usetikzlibrary{positioning}
\usepackage[pscoord]{eso-pic}
\usepackage{framed}
\usepackage{multirow}

\usepackage{etoolbox}
\usepackage{framed}
\newcommand{\tmpcomment}[3]{{\emph{\textcolor{#1}{#2: #3}}}}

\newtoggle{hidecomments}
\toggletrue{hidecomments}

\iftoggle{hidecomments}{
	\renewcommand{\tmpcomment}[3]{}
}{
}

\usepackage{url} 

\newcommand{\accessedDate}{April~2021}
\newcommand{\urlfootnote}[1]{\footnote{\url{#1} Accessed \accessedDate}}

\newcommand{\placetextbox}[3]{
  \AddToShipoutPictureFG*{
    \put(\LenToUnit{#1\paperwidth},\LenToUnit{#2\paperheight}){\vtop{{\null}
        \framebox[\textwidth]{\parbox{\dimexpr\textwidth-2\fboxsep-2\fboxrule}{\footnotesize{#3}}}  
    
    }}%
  }%
}%

\def\BibTeX{{\rm B\kern-.05em{\sc i\kern-.025em b}\kern-.08em
    T\kern-.1667em\lower.7ex\hbox{E}\kern-.125emX}}
\begin{document}

\title{Speech Quality Assessment in Crowdsourcing: Comparison Category Rating Method}
\author{
    \IEEEauthorblockN{Babak Naderi\IEEEauthorrefmark{1}, Sebastian M{\"o}ller\IEEEauthorrefmark{1}\IEEEauthorrefmark{2},
    Ross Cutler\IEEEauthorrefmark{3}}
    \IEEEauthorblockA{\IEEEauthorrefmark{1}Quality and Usability Lab, Technische Universit{\"a}t Berlin, Germany, forename.surname@tu-berlin.de}
    \IEEEauthorblockA{\IEEEauthorrefmark{2}DFKI Projektb{\"u}ro Berlin, Germany, sebastian.moeller@dfki.de}
    \IEEEauthorblockA{\IEEEauthorrefmark{3}Microsoft Corp., USA, ross.cutler@microsoft.com}
} 



\maketitle

\begin{abstract}
Traditionally, Quality of Experience (QoE) for a communication system is evaluated through a subjective test. The most common test method for speech QoE is the Absolute Category Rating (ACR), in which participants listen to a set of stimuli, processed by the underlying test conditions, and rate their perceived quality for each stimulus on a specific scale. The Comparison Category Rating (CCR) is another standard approach in which participants listen to both reference and processed stimuli and rate their quality compared to the other one. The CCR method is particularly suitable for systems that improve the quality of input speech.
This paper evaluates an adaptation of the CCR test procedure for assessing speech quality in the crowdsourcing set-up. The CCR method was introduced in the ITU-T Rec. P.800 for laboratory-based experiments. We adapted the test for the crowdsourcing approach following the guidelines from ITU-T Rec. P.800 and P.808. We show that the results of the CCR procedure via crowdsourcing are highly reproducible. We also compared the CCR test results with widely used ACR test procedures obtained in the laboratory and crowdsourcing. Our results show that the CCR procedure in crowdsourcing is a reliable and valid test method.
\end{abstract}

\begin{IEEEkeywords}
crowdsourcing, speech, quality assessment, comparison category rating, CCR
\end{IEEEkeywords}

\section{Introduction}
\placetextbox{0.08}{0.08}{©2021 IEEE. Personal use of this material is permitted. Permission from IEEE must be obtained for all other uses, in any current or future media, including reprinting/republishing this material for advertising or promotional purposes, creating new collective works, for resale or redistribution to servers or lists, or reuse of any copyrighted component of this work in other works. This paper has been accepted for publication in the 2021 Thirteen International Conference on Quality of Multimedia Experience (QoMEX).}

Speech quality is a common object of assessment for telecommunication service providers in order to optimize the quality experienced by their customers. As it is subjective in nature, speech quality can only be assessed using subjective listening-only or conversation test procedures. For the listening situation, which is the focus of this paper, ITU-T Rec. P.800~\cite{ITU-P800} recommends different test methods, including Absolute Category Rating (ACR), Degradation Category Rating (DCR) and Comparison Category Rating (CCR). Whereas ACR is better suited to obtain a general, unbiased judgment of the overall quality, DCR and CCR might be better suited for smaller, subtle differences~\cite{ITU-P800}. In all these methods, short speech stimuli (4-8 s) are presented to listening participants in a controlled, quiet lab environment.

In ACR, stimuli are presented in isolation and test participants are asked to rate the overall quality of each stimulus on a 5-point discreet scale, ranging from ``bad'' = 1 to ``excellent'' = 5. For each stimulus and/or degradation condition, average scores are computed, resulting in a Mean Opinion Score (MOS). The MOS values should be accompanied with basic information about the distribution of ratings like number of votes and standard deviation or the 95\% Confidence Interval (CI). In DCR and CCR, two stimuli with the same linguistic content are presented to the participants, one containing the (clean speech) \textit{reference} and another containing a \textit{processed} version (i.e. the reference stimulus processed by the system under the test). In DCR, the reference stimulus is presented first, and it is the task of the participants to rate the second, processed stimulus on a five-point scale from ``degradation is inaudible'' = 5 to ``degradation is very annoying'' = 1. In CCR, participants do not know which signal is the reference and which is the processed one, as their order is randomized. The participants' task is to rate the quality of the second stimulus compared to the first one on a scale from ``much worse'' = -3 to ``much better'' = +3. The resulting average scores are called DMOS and CMOS, respectively. Both methods are particularly useful for assessing the communication system when the input has been degraded by background noise~\cite{ITU-P800}. The advantage of CCR procedure over the DCR method is the possibility to assess speech system that either degrade or enhance the quality of the speech (e.g. noise cancellation systems)~\cite{ITU-P800}. The main disadvantage of both methods compare to the ACR is that the test duration will be longer as the test participant should listen to both reference and processed stimuli.

Whereas the standard ACR, DCR and CCR procedures are all carried out in a controlled lab environment to reduce the impact of variables extraneous to the purpose of the test, a rather recent strategy has been to carry out quality assessment tasks in an anonymous crowd of internet users, on a microtask platform on which participants are paid for their service. 
Consequently, crowdsourcing offers major benefits for subjective testing, including lower costs, higher speed, more flexibility, scalability, and access to a diverse group of participants~\cite{hossfeld2014best} . In turn, the listening happens in the participants' natural (home or work) environment~\cite{jimenez2019background}, under less controlled conditions which might include ambient noise, non-calibrated device, non-attention due to parallel tasks, or non-cooperative participants~\cite{naderi2020towards,naderi2018motivation}. In order to reduce the impact of these variables on the obtained test results, ITU-T Rec. P.808 \cite{ITU-P808} provides recommendations on how to perform the speech quality assessment using crowdsourcing, however so far limited to the ACR paradigm. In addition, an open-source P.808 Toolkit \cite{naderi2020open} has been developed which includes an implementation of the DCR and CCR test methods according to the ITU-T Rec. P.800 \cite{ITU-P800}, adapted to some requirements of ITU-T Rec. P.808. Still, the reliability and validity of using CCR and DCR in the crowd have not been analyzed yet.

In this paper we compare the results of multiple CCR crowdsourcing test to assess its reproducibility. Furthermore, we also compare the result of the CCR crowdsourcing test to the results of two ACR tests, one carried out in the laboratory (according to the ITU-T Rec. P.800) and one in the crowd (according to the ITU-T Rec. P.808). Section II explains the implementation of CCR test tailored to the crowdsourcing and presents two experiments we conducted. In section III results of the experiments are presented. In section IV we provide a comparison between MOS and CMOS. A discussion concludes the paper in Section V.

\section{Method}
We have conducted two types of experiments. In the first experiment, we compared the CCR implementation in the P.808 Toolkit with ACR ratings in the laboratory and crowdsourcing. In the second experiment, we enhanced the implementation and compared the result of multiple CCR tests in crowdsourcing to check the reliability of the test method.

\subsection{CCR test procedure for Crowdsourcing}
\begin{figure}[htbp] 
   \centering
   \begin{subfigure}[b]{0.5\textwidth}
     \centering
     \includegraphics[width=\textwidth]{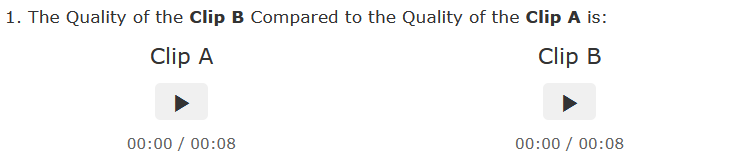} 
     \caption{} 
     \label{figd:a} 
   \end{subfigure} 
   ~
   \begin{subfigure}[b]{0.5\textwidth}
     \centering
     \includegraphics[width=\textwidth]{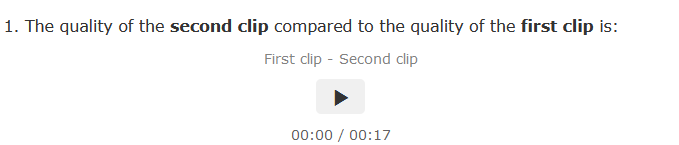}
     \caption{}
 	\label{figd:b} 
   \end{subfigure} 
   ~
   \caption{Two designs of the stimuli presentation for CCR test (a)~Experiment I and (b)~Experiment II.}
   \label{fig:design} 
 \end{figure}

Like the crowdsourcing ACR test, in the CCR test session, the eligibility of test participants and suitability of their environment and setup should be tested remotely. The crowdsourcing test session has 5 sections: \textit{Instruction}, \textit{Qualification}, \textit{Setup test}, \textit{Training}, and \textit{Rating}. The \textit{qualification} section includes a hearing test (digit-triplet test~\cite{levitt1992adaptive}) besides the typically demographic questions, which is only shown once. In case there is a need to evaluate participants' language proficiency, a corresponding test should be added to this section. The \textit{setup} section includes listening level adjustment, headset usage test, and environment test (Modified Just-Noticeable Difference in Quality test \cite{naderi2020env}). 
The session also contains a periodical \textit{training} section \cite{polzehl2015robustness} to familiarize the participants with the test procedure and presenting the anchoring test stimuli. The participants are exposed to the training section only when 60 minutes or more is passed from the last training. 
Finally, in the \textit{rating} section 10 to 12 stimuli are presented to the participant for rating, including a gold standard question. 
We designed the gold standard questions for the CCR in the way that both stimuli presented to participants are the reference signals. Therefore, we expect the user rate \textit{“About the Same”}=0 in the CCR scale.
Within the training section, we also included a gold standard question which in case participant's response is not as expected (i.e. About the same), a feedback message is shown. 
We also used two different user interfaces (c.f. Figure \ref{fig:design}). In Experiment I, we used the P.808 Toolkit implementation in which two Audio Playback components are shown per question (one for Clip A and the other one for Clip B). The participant can play them in arbitrary order, and after listening to both, then can rate the quality of Clip B compared to the Quality of Clip A on the CCR scale (Figure \ref{figd:a}).
However, we noticed in some cases participants, who listened to both audio clips multiple times, mistakenly answered to the question the other way around i.e. rated the quality of the last clip they heard compare to the other one. Therefore we created the second implementation.

In the second implementation (Experiment II), only one Audio Playback component was shown on the page, which plays the first and second clips after each other with one-second silence in between. The component shows which audio clip is currently playing. After listening to both stimuli,  participants are asked to cast their vote on the second stimulus’s quality compared to the first stimulus they had listened to on the CCR scale (Figure \ref{figd:b}).
We randomize the order of stimuli (i.e., referenced and processed one) so that about half of the participants listen to the referenced clips first and the other half to the processed one. 
Consequently, we corrected the collected votes before calculating the CMOS in a way that it always represents the quality of processed signal compared to the reference one (see section E.5 of ITU Rec. P.800 for more details).

\subsection{Experiment I: Comparison between CCR test in crowdsourcing and ACR test}
For this experiment, we used dataset 401 from the pool of the ITU-T Rec. P.863 competition datasets \cite{ITU-P863}.
The dataset contains 48 degradation conditions and 24 stimuli per condition. Results of the ACR test in the standard laboratory room, based on the ITU-T Rec. P.800, were kindly provided to us. We calculated the MOS values based on the 192 votes per test conditions (hereafter MOS Lab). Meanwhile, this dataset has previously been used in an ACR crowdsourcing test as well \cite{naderi2020towards, naderi2020impact} in which authors made the ratings openly available. We calculated the MOS values using on average 217 valid votes per test condition  (hereafter MOS CS). We also conducted a crowdsourcing experiment following the CCR method in the Amazon Mechanical Turk (AMT)\urlfootnote{www.mturk.com}. In this study, we used the open-source P.808 Toolkit \cite{naderi2020open}. The rating section contained ten stimuli and one gold stimulus as explained before. We aimed to collect nine votes per stimulus. In all crowdsourcing studies reported in this paper, we recruited workers from US who has more than 500 tasks accepted with more than 98\% acceptance rate in all their tasks across job providers\footnote{These filters are freely available in AMT, which help to filter participants based on their location and quality and quantity of their previous work.}. 

\subsection{Experiment II: Reproducibility CCR test in crowdsourcing}
We used an openly available dataset from Experiment~2 of the ITU-T Supplement  23\cite{ITU-Psup23} which was designed to evaluate the Terms of Reference for codec performance under conditions of environmental background noise and background music using the CCR test method. The dataset contains 40 degradation conditions and 136 stimuli\footnote{For some of the conditions four clips and for the reference conditions only two clips are included.} in American English. The dataset contains two codecs i.e. G729 and G726, each with one and two encodings and seven conditions of background noise i.e. clear or with different types and levels of background noises (2 codecs x 2 encodings x 7 noise conditions). The rest of conditions belong to the reference conditions which can be divide in two groups: 1) five conditions in which the signal was degraded using Modulated Noise Reference Unit (i.e. MNRU from 6~dB to 30~dB) and 2) seven conditions which not processed by coded (i.e. Direct) but included different type and level of background noise (including a clear condition with no noise). Further details on the dataset can be found in \cite{ITU-Psup23}.  
For this dataset, results of a CCR subjective test in the laboratory, which was conducted according to the ITU-T Rec. P.800 at the beginning of 1990s, are also openly published by the ITU-T (hereafter CMOS Lab) \cite{ITU-Psup23}. It includes 24 subjective ratings per stimuli.

Using this dataset, we conducted three crowdsourcing CCR tests with different groups of workers in different days, using our enhanced implementation (i.e. Figure~\ref{figd:b}). In each experiment we aimed to collect 30 votes per stimuli pair.

\section{Results}

\subsection{Experiment I:}
In the CCR crowdsourcing study, from 1044 submissions, 726 passed the data screening step (used both earpods, correct answer to the gold stimulus, and passed the environment test\cite{naderi2020env}). On average, we collected 151 accepted votes per condition (STD = 7.7). We calculated the CMOS in a way that -3 shows the lowest quality and 0 shows the highest quality (as explained in section E.5 of ITU Rec. P.800).

\begin{figure*}[htb] 
   \centering
   \begin{subfigure}[b]{0.3\textwidth}
     \centering
     \includegraphics[width=\textwidth]{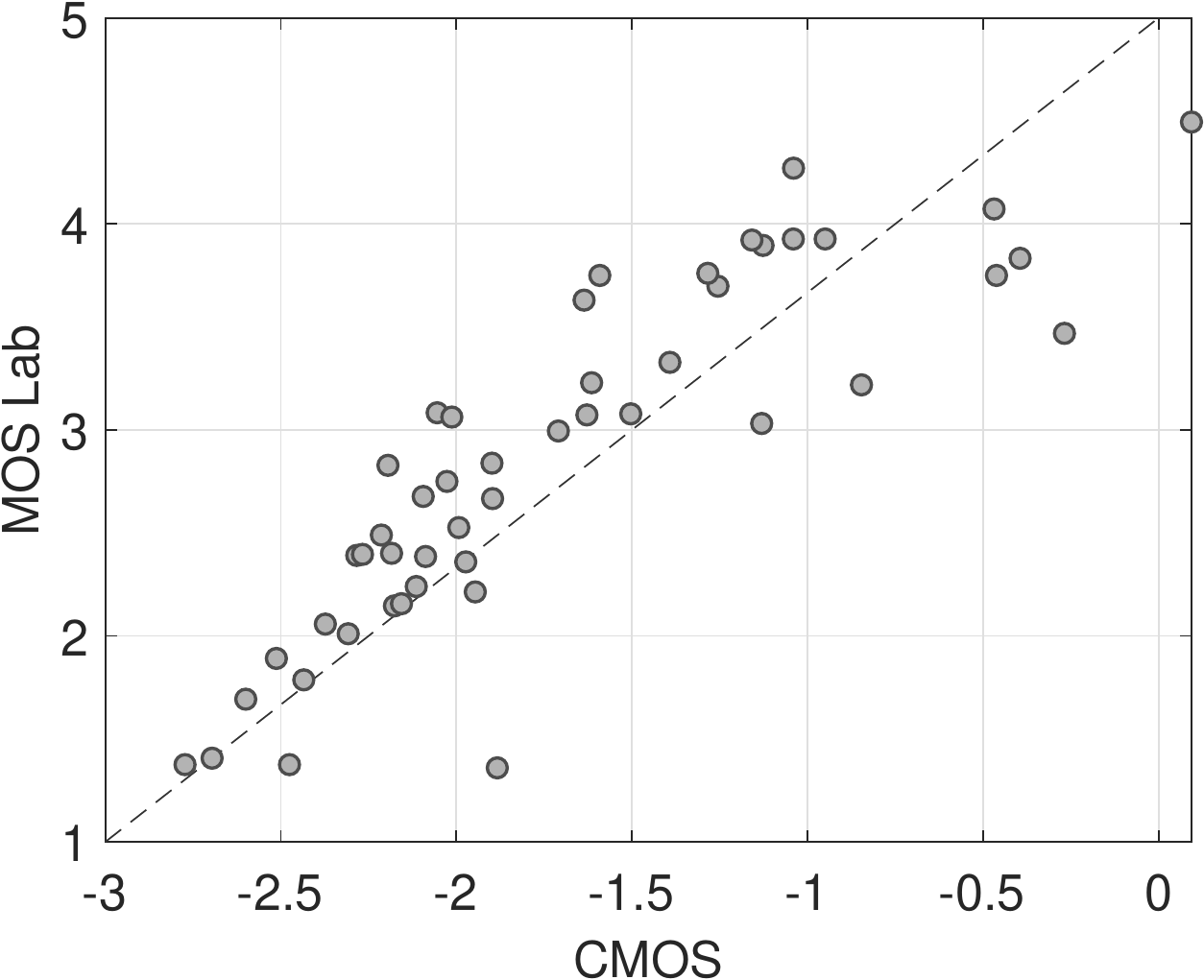} 
     \caption{} 
     \label{figcm:a} 
   \end{subfigure} 
   ~
   \begin{subfigure}[b]{0.3\textwidth}
     \centering
     \includegraphics[width=\textwidth]{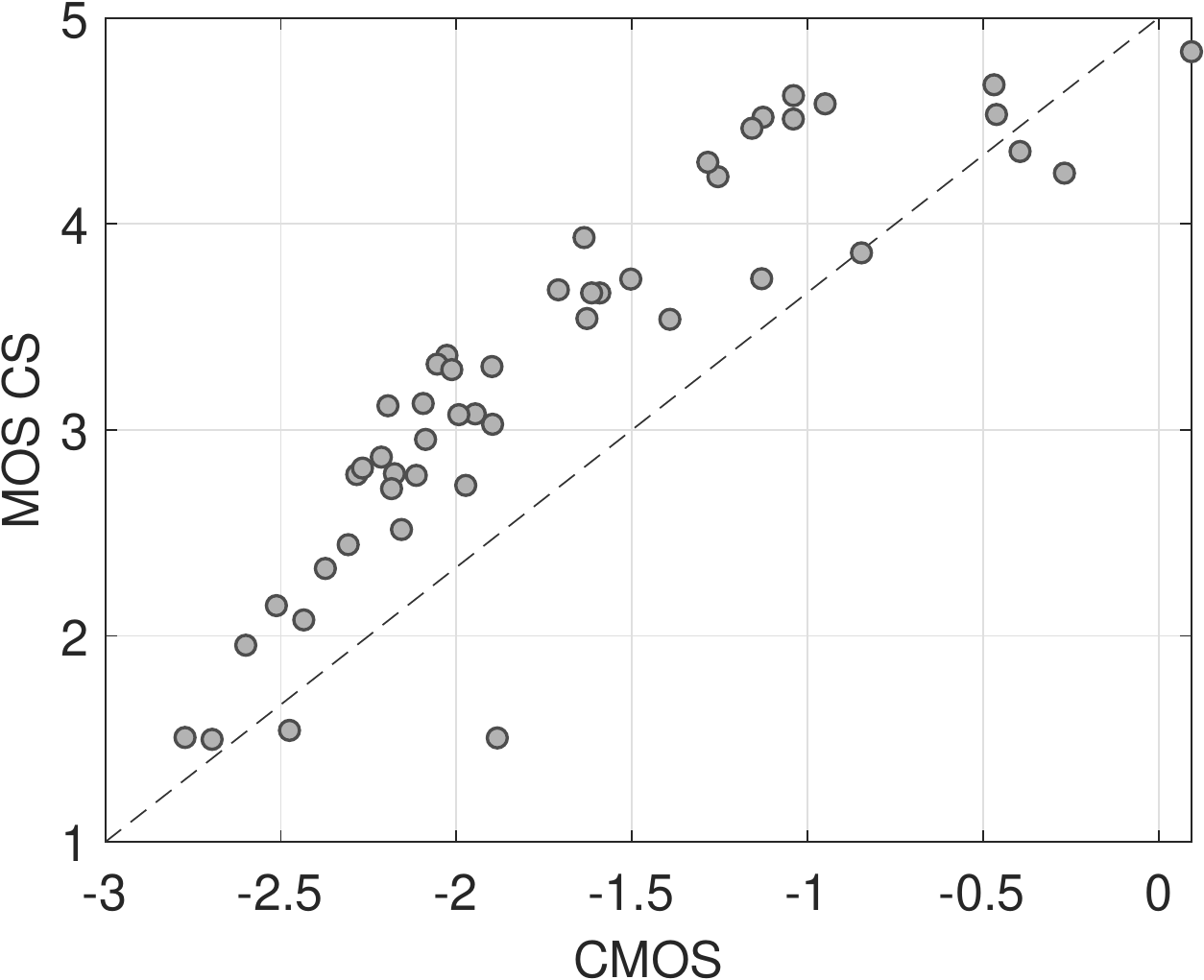}
     \caption{}
 	\label{figcm:b} 
   \end{subfigure} 
   ~
  \begin{subfigure}[b]{0.3\textwidth}
       \centering
     \includegraphics[width=\textwidth]{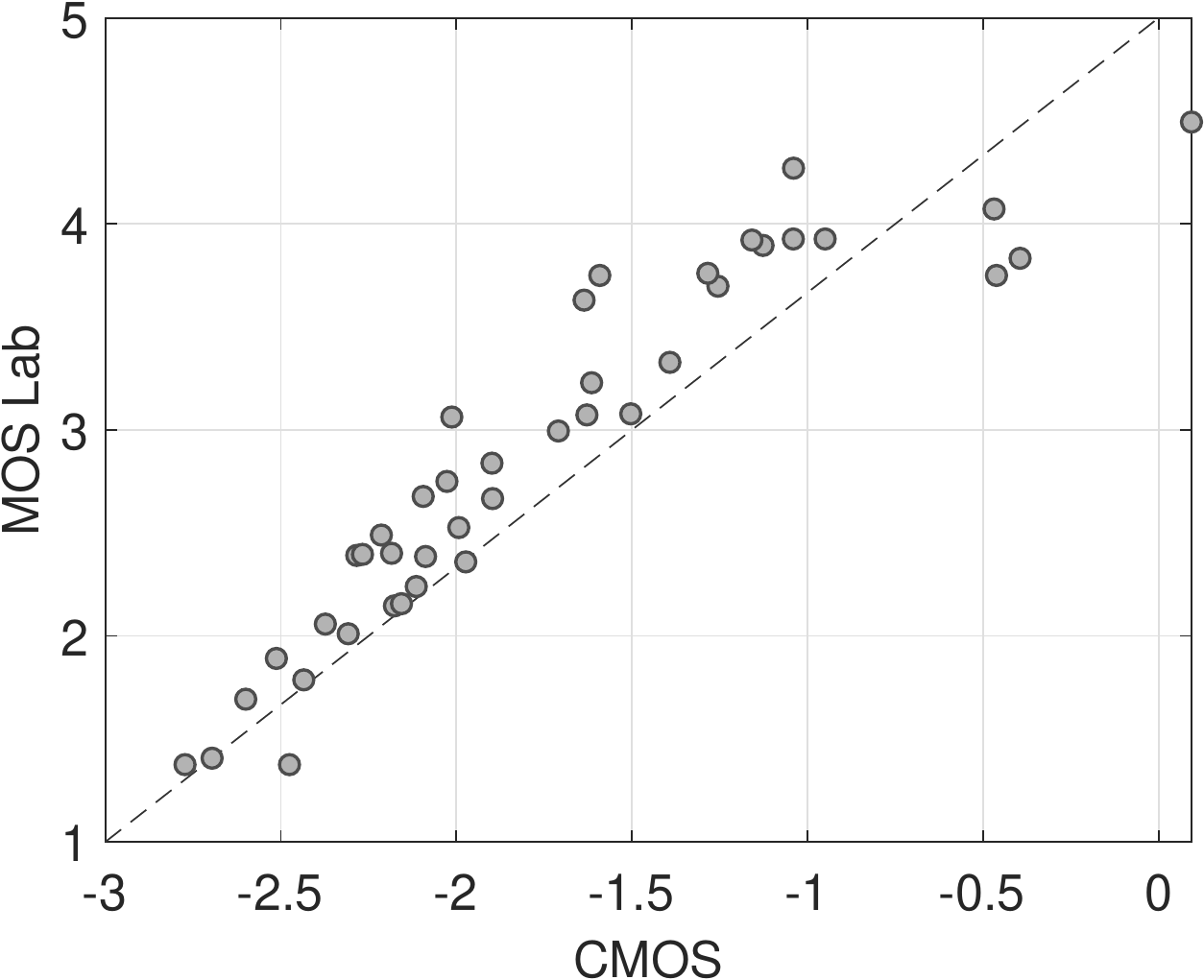}
     \caption{}
 	\label{figcm:c} 
   \end{subfigure} 
    ~
     \begin{subfigure}[b]{0.3\textwidth}
     \centering
     \includegraphics[width=\textwidth]{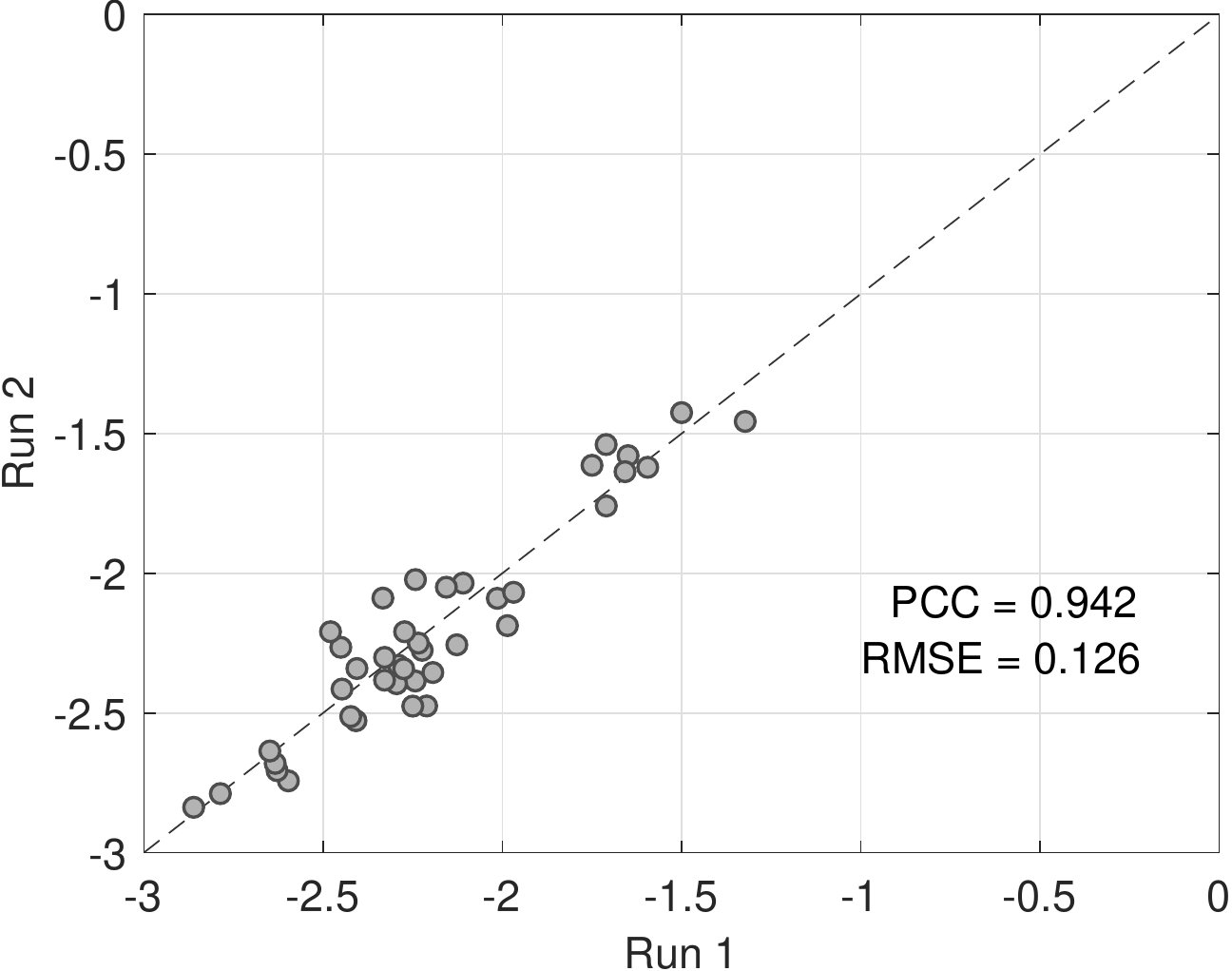} 
     \caption{} 
     \label{figcm:d} 
   \end{subfigure} 
   ~
   \begin{subfigure}[b]{0.3\textwidth}
     \centering
     \includegraphics[width=\textwidth]{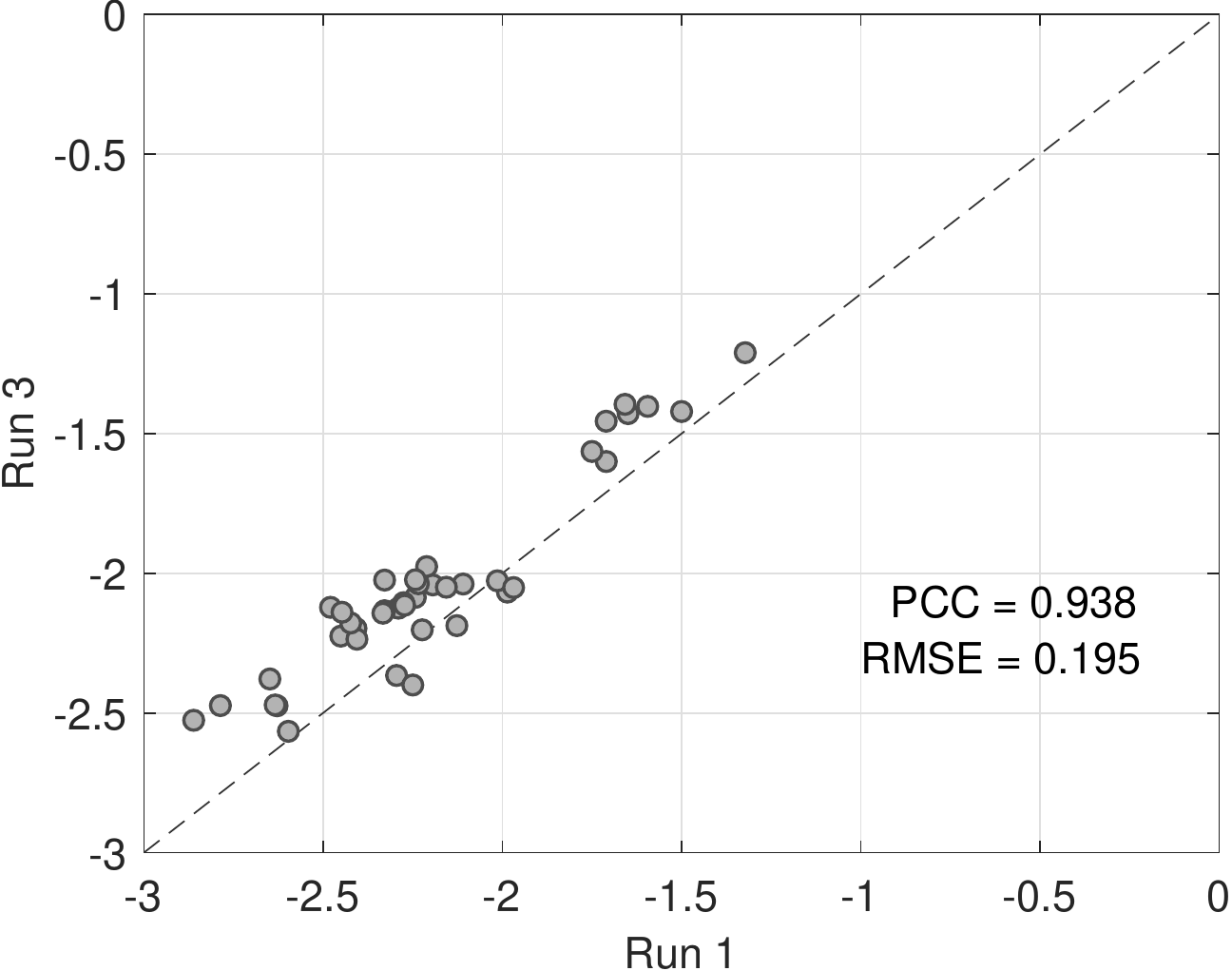}
     \caption{}
 	\label{figcm:e} 
   \end{subfigure} 
   ~
  \begin{subfigure}[b]{0.3\textwidth}
       \centering
     \includegraphics[width=\textwidth]{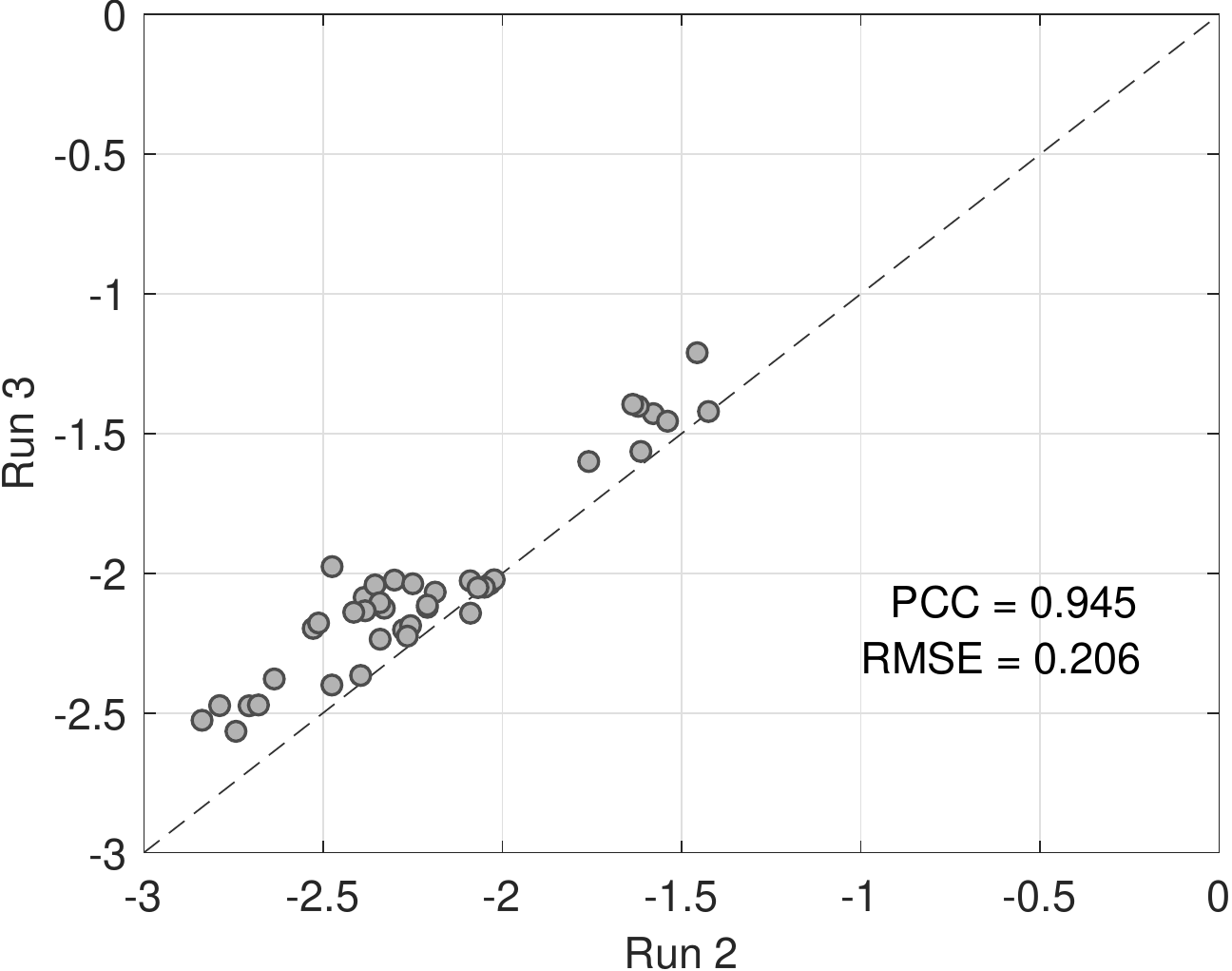}
     \caption{}
 	\label{figcm:f} 
   \end{subfigure} 
   \caption{(a-c) CMOS compare to the MOS from (a) Lab and (b) crowdsourcing experiments. (c) CMOS vs Lab Mos when outliers (Table \ref{tab:study1_outlier}) are removed. (e-f) Distribution of CMOS values in three runs in reproducibility experiment.}
   \label{fig:result:cmos_mos} 
 \end{figure*}

Pearson correlation coefficient (PCC) and Spearman’s ranked correlation coefficient (SRCC) between CMOS, MOS Lab and MOS CS are reported in Table \ref{tab:study1_corr}. Figure \ref{figcm:a}, \ref{figcm:b} also illustrates the distributions of MOS Lab and MOS CS compare to the CMOS values in scatter plots.

Further investigations showed that the position of seven conditions in the rank order of CMOS strongly differ from their positions in the rank order of MOS Lab. Mostly, their quality is rated higher in CCR. Those conditions are reported in Table \ref{tab:study1_outlier}. By removing all of the seven conditions, the PCC between CMOS and MOS increased to 0.96 for both lab and CS. After removing all of the seven conditions, the resulting scatter plot is illustrated in Figure \ref{figcm:c}. It should be noted that the CMOS is expected to have a range of -3 to 0 (4 points), whereas MOS has a range of 1 to 5 (5 points).

\begin{table}[hbt]
\caption{Pearson correlation coefficient (upper-triangle) and Spearman’s rank correlation coefficient (lower-triangle) between CMOS and ACR MOS from lab and CS tests (Study I).}
\label{tab:study1_corr} 
\begin{center}
\begin{tabular}{l c c c}
\toprule
& \textbf{CMOS CS} & \textbf{MOS Lab} &  \textbf{MOS CS}\\
\midrule
\textbf{CMOS CS}     &       & .862 & .883\\
\textbf{MOS Lab}  &.877   &       & .98\\
\textbf{MOS CS}  & .913  & .971  & \\
\bottomrule
\end{tabular}
\end{center}
\end{table}

\begin{table}[hbt]
\caption{Conditions with strong deviation of their position in rank order of MOS Lab and CMOS (Experiment I). PL: Packet loss, PF: Profile}
\label{tab:study1_outlier} 
\begin{center}
\resizebox{\columnwidth}{!}{%
\begin{tabular}{p{4mm}ccccc  l}
\toprule
\textbf{Cond.} & \multicolumn{3}{c}{\textbf{Rank}}  & \textbf{CMOS} & \textbf{MOS} & \textbf{Description}\\
& \textbf{CMOS} & \textbf{MOS} & \textbf{Delta} & & & \\
\midrule
C09 & 37 & 25 & 12 & -2.19 & 2.83 & Band pass 500-2500 Hz (NB)\\
C11 & 2 & 14 & -12 & -0.27 & 3.47 & Temp. clipping - 2\% PL (SWB)\\
C12 & 22 & 48 & -26 & -1.88 & 1.36 & Temp. clipping - 20\% PL (SWB)\\
C18 & 25 & 37 & -12 & -1.95 & 2.21 & VoIP client 2-loss PF3 (WB)\\
C34 & 30 & 18 & 12 & -2.05 & 3.08 & VoIP client 1-loss PF1-to Mobile (NB)\\
C45 & 6 & 17 & -11 & -0.85 & 3.22 & G.722.1 Annex C-loss PF2 (SWB)\\
C46 & 11 & 22 & -11 & -1.13 & 3.02 & G.722.1 Annex C-loss PF3 (SWB)\\

\bottomrule
\end{tabular}
}
\end{center}
\end{table}

\subsection{Experiment II}
On average, 62.85\% of submissions in the three crowdsourcing tests passed the data cleansing step (used both earpods, correct answer to gold stimulus, passed the environment test).
Consequently, we have collected on average 60.1, 69.3 and 66.2 valid votes per test conditions in run1-3, respectively. We calculated the CMOS values and the 95\%CI per test condition per run.
The average 95\%CIs were in the range of [0.178, 0.198] for the three runs.
The PCC and SRCC between CMOS values of three runs are reported in Table~\ref{tab:study2_corr}. 
The result shows a high PCC between the three runs and consequently high test-retest reliability.

Figure~\ref{figcm:d}--\ref{figcm:f} illustrate the distribution of CMOS values between different runs. An offset between Run3 and the others is visible. After applying a first-order mapping on the CMOS values from Run3, the average RMSE between three studies drops from 0.175 to 0.125.
One possible reason for the smaller SPCC, is that all conditions have very close CMOS values i.e. the average range between CMOS of the worse and the best conditions in the three studies was 1.436 CMOS.

We fitted a linear mixed-effects model (LMEM) with random intercepts and with the \textit{run} and the \textit{degradation condition} as fixed factors and \textit{participant} as random factor. Result shows there is only a significant main effect from degradation condition $F(39,7740.4)$ = $35.694$, $p<0.001$. Consequently neither a significant main effect from run nor an interaction effect was observed.

We also compared our results with the openly available laboratory results published in the ITU-T Sup23, reaching PCC = 0.17. A close investigation of reference conditions (C34-40 with direct codec) revealed that the CMOS values from the laboratory test were not influenced by the type and level of background noise impairments. Although the reference conditions contain Office Babble, Street, Hoth, White, and Music background noise at 20 dB SNR and Vehicular at 10 and 20 dB SNR, their quality were judged to be as low as -0.13 CMOS only.
As a result of further investigation (and consulting with delegates involved in conducting the laboratory experiments), we hypothesis that in the laboratory test one-ear listening condition, high environmental noise in the test room (due to hardware MNRU module) and participants expectation of telephony call in 1990s should be the reason of them neglecting the effect of background noise in the speech files when rating them.
Finally, we conclude that the CMOS Lab results reported in ITU-T Sup.23 should not be considered as ground truth and reproducible anymore\footnote{Note that 75\% of conditions in Experiment2 of the ITU-T Sup23 at least partly degraded with background noise.}.

Furthermore, a closer look at the test conditions shows that the obtained CMOS values in our crowdsourcing studies are inline with the levels of impairments in the processed stimuli i.e. increase in the SNR level or the ratio of speech power to modulated noise power in MNRU conditions lead to a better quality rating, as expected. 

In addition, we compared the three runs to check if they lead to the same set of conclusions. Here, the conclusion means either two degradation conditions are significantly different or not. As the dataset was built to evaluate the performance of codecs in conditions with background noise, we conducted two-way Analysis of Variance (ANOVA) to compare the overall quality achieved by two different codecs on the seven different background noise conditions. As the dataset contains one- and two-times coding, we conducted two ANOVAs per run. 
The ANOVAs yielded to significant main effects of background noise and no interaction effect between noise and codec in all three runs. 
Detailed paired comparison of the main effect between different noise conditions with Bonferroni correction (21 combinations), revealed that in 71\% cases with one-time coding and 85.7\% cases in two-times coding, all three runs reach the same conclusions. 
Most of discrepancy happens when comparing 10 SNR\textsubscript{db} of Vehicle background noise with other background noises (beside Music) in one-time coding. 

Finally, We calculate the Intra-class Correlation Coefficient (ICC) to further examine the reliability of the CCR implementation based on a single measure, absolute agreement and two-way mixed random model i.e. ICC(A,1). It demonstrates how strongly each run of the CCR implementation resembles each other. The Result shows excellent reliability (ICC= 0.94).

\begin{table}[hbt]
\caption{Pearson correlation coefficient (upper-triangle) and Spearman’s rank correlation coefficient (lower-triangle) between CMOS of three runs in crowdsourcing.}
\label{tab:study2_corr} 
\begin{center}
\begin{tabular}{c c c c}
\toprule
& \textbf{CMOS Run1} & \textbf{CMOS Run2} &  \textbf{CMOS Run3}\\
\midrule
\textbf{CMOS Run1}  &       & .942  & .938 \\
\textbf{CMOS Run2}  &.848   &       & .945 \\
\textbf{CMOS Run3}  &.86   & .841   &  \\
\bottomrule
\end{tabular}
\end{center}
\end{table}

\section{Comparison between ACR and CCR}

In this section, we compare the ACR and CCR methods from different perspectives:
\subsubsection{Method} Although the CCR scale has 7 points, but mostly one side of the scale is relevant i.e. either form -3 to 0 or 0 to 3 depending to the system under the study. Consequently quality scores in CCR are distributed in smaller range (4 points) than the ACR scale (5 points). This leads to a banana-shaped distribution when CMOS compared to the MOS in Figure~\ref{figcm:a}-\ref{figcm:b}. Consequently the CCR scale has a lower discriminating power than ACR scale. Meanwhile in the CCR test participants listen to both reference and processed stimuli which leads to a longer test session and higher costs (roughly $\times2$).

\subsubsection{Bias or offset} Similar to the ACR test, subjective ratings from CCR tests in multiple studies may differ by an offset and/or gradient. This effect is well-known for the ACR test~\cite{ITU-P1401} and was observed in Experiment~2 for the CCR test.

\subsubsection{Uncertainty} 
\begin{figure}[htb]
 \centering
 \includegraphics[width=0.8\columnwidth]{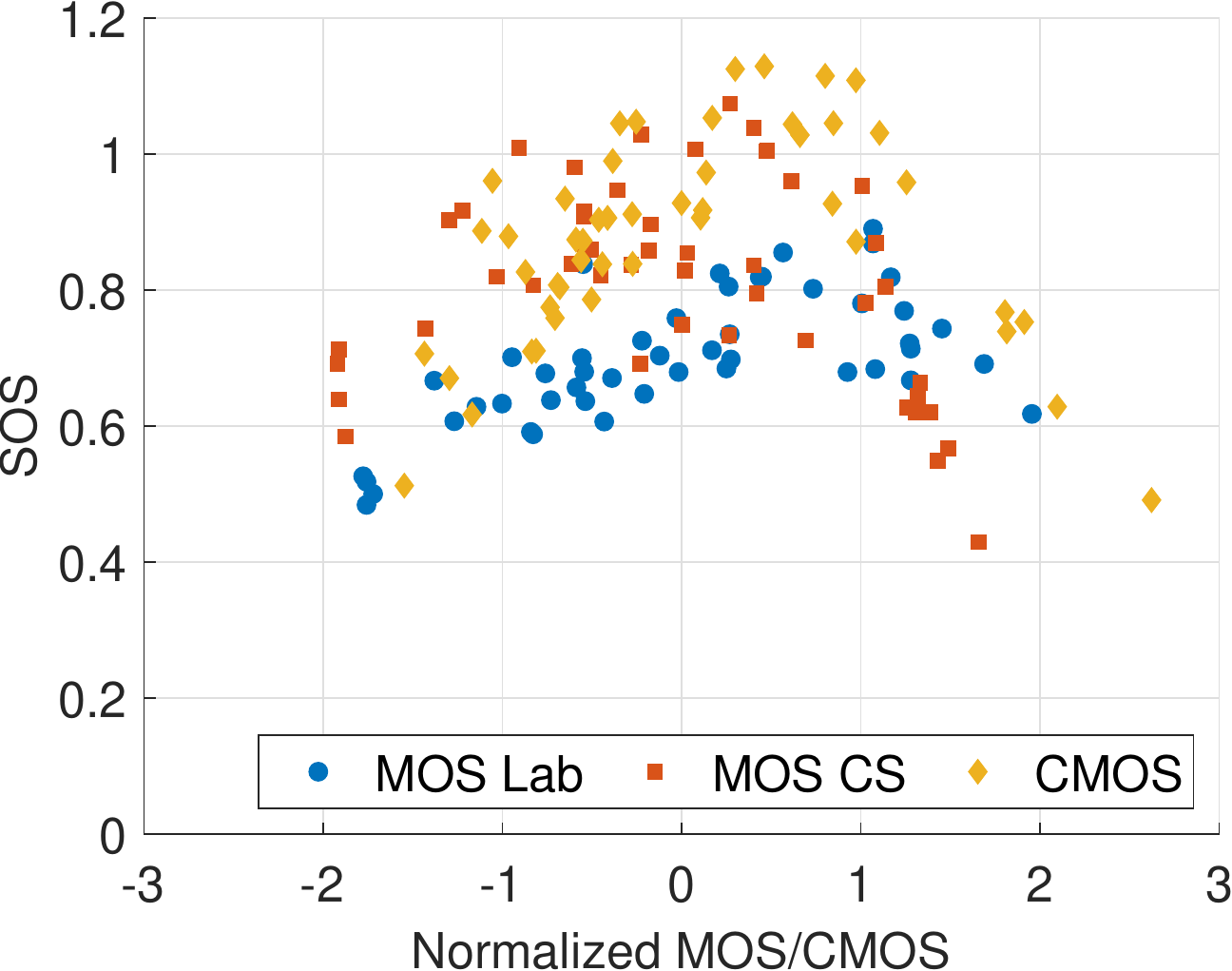}
 \caption{SOS-normalized Mean Scores plot for different subjective tests in the Experiment I.}
 \label{fig:results:sosmos}
\end{figure}

Raters disagreement or standard deviation of ratings are higher in the crowdsourcing than the laboratory but no difference was observed between CCR and ACR. Figure~\ref{fig:results:sosmos} illustrate the distribution of SOS~\cite{hobetafeld2011sos}(i.e. standard deviation of mean score) over the normalized mean scores from MOS Lab, MOS CS and CMOS in Experiment I.

\subsubsection{Impairments}
Considering various impairments in Experiment I, we calculated the difference between each test condition's position in the rank-order of the CMOS and the MOS Lab.
We also used an instrumental DNN model~\cite{mittag2021diss} to predict the speech quality dimensions (i.e. Noisiness, Discontinuity, Coloration and Loudness \cite{waltermann2013dimension}) per each clip and aggregated them per test condition for Experiment I. The DNN model has shown excellent performance on predicting subjective ratings on the quality dimensions in ~\cite{mittag2021diss} with average RMSE\textsuperscript{*}=0.2 for all four dimensions.

We observed a significant PCC= 0.662, $p< .001$ between the test condition's discontinuity score and the difference between its rank-order in CMOS and MOS Lab\footnote{No other significant correlation with the rest of quality dimensions was observed.}. In other words, participants in the CCR test rated the test conditions with discontinuity impairment to have a better quality (relative to other test conditions) compare to the test participants in the ACR test. We hypothesize that as the CCR test participants listened to the reference signal before or after the processed signal, they perceived discontinuity less disturbing as they unconsciously got the content. Therefore, they rated the discontinuity not as bad as participants in an ACR test.
From the conditions reported in Table~\ref{tab:study1_outlier} which have a substantial deviation of their rank order between two tests, five include the discontinuity confirmed by an expert review (i.e. C11, C12, C18, C45 and C46). From them, C12 is strongly and only impaired by discontinuity. 

In addition, test conditions that include coloration impairments have got a lower rank in the CCR. We believe as the listener could compare the voice in the processed and reference samples and recognize how pure the voice remained in the process. The coloration was the dominant impairment in conditions C9 and C34 from Table~\ref{tab:study1_outlier}.

\section{Discussion and Conclusion}
In this paper, we showed that the CCR test method in crowdsourcing produces reliable and valid results. We showed that the obtained CMOS values align with the levels of impairments in the processed stimuli and the test method provides highly reproducible results.

We also compared the results of ACR and CCR test methods. The CCR test method is more costly as participants listen to both reference and the processed stimuli before voting. The CCR scale also has a lower discriminating power as only one side of the scale is typically relevant. Both ACR and CCR showed a similar distribution of uncertainty in terms of standard-deviation of crowdsourcing ratings. 
Impairments affecting discontinuity and coloration of speech samples were perceived differently in the CCR test than ACR. Participants in the CCR test perceived test conditions with discontinuity impairment to have a higher quality than their counterparts in the ACR test. Contrariwise they rated test conditions with coloration impairment having lower quality. We did not observe a significant difference between test conditions' judgment impaired by noisiness or loudness between two test methods.

One limitation of this comparison is the dataset used for comparing ACR and CCR. The dataset used in Experiment~1 previously used for standardization activities and covered various degradation conditions, which facilitates comparing two test methods for degradations affecting all four quality dimensions. However, a dataset that includes nuanced background noise differences could better demonstrate the advantages of the CCR method. That should be a subject of future work. Beside that, for future work, the performance of the CCR test for evaluating speech enhancement algorithms shall be investigated. Listening to both the reference and the processed signals should facilitate recognizing smaller improvements. Meanwhile, the suitability of other scales (e.g. 9-point or continuous scales) for the CCR method should be evaluated to address the scale's discriminating power.
Finally, we made the ratings collected in the CCR tests publicly available for further investigation\urlfootnote{https://github.com/babaknaderi/Crowdsourced-speech-quality-CCR-test}. Our enhancement is also applied in the P808 Toolkit and are openly available which can be used for conducting CCR speech quality test in the crowdsourcing.


\bibliographystyle{IEEEtran}
\bibliography{lib}

\end{document}